
\tolerance=10000

\def\np{Nucl. Phys.}

\input phyzzx.tex

\def\Tr{{\rm Tr}}

\Pubnum = {QMW-93-25}
\date = {October 1993}
\titlepage
\title {\bf  GAUGED HETEROTIC SIGMA-MODELS}
\author {C. M. Hull}
\
\address {Physics Department,
Queen Mary and Westfield College,
\break
Mile End Road, London E1 4NS, United Kingdom.}
\abstract
{The gauging of isometries   in general sigma-models which include fermionic
terms which represent the interaction of strings with background Yang-Mills
fields is considered. Gauging is possible only if certain obstructions are
absent. The quantum gauge anomaly is discussed, and the (1,0) supersymmetric
generalisation of the gauged action given.}

\endpage
\pagenumber=1

\REF\sen {A. Sen, Phys. Rev. D3 (1985) 2162; Phys Rev Lett 55
         (1985) 1846;  C. Callan, D. Friedan,
                  E. Martinec and M.J. Perry, \np 262
            (1985 593.}
\REF\gauged{             K. Bardacki,
              E. Rabinovici and B. S\"aring, \np 299 (1988) 157;\hfill\break
             H.J. Schnitzer, \np 324 (1989) 412;\hfill\break
             D. Karabali and H.J. Schnitzer, \np 329 (1990) 649;\hfill\break
             K. Gaw\c edzki and A. Kupiainen, Phys. Lett. 215B (1988) 119,
             Nucl. Phys. B320(FS) (1989) 625.}
\REF\rocver {M. Ro\v cek and E. Verlinde, Nucl. Phys.
{\bf B373} (1992), 630.}
 \REF\queossa {X.C. de la
Ossa and F. Quevedo, Nucl. Phys. {\bf B403} (1993), 377.}
\REF\pots{C.M. Hull, G. Papadopoulos and P.K. Townsend, {\it Potentials for
(p,0) and
(1,1) supersymmetric sigma-models with torsion}, preprint DAMTP/R-93/8.}
\REF\hso{C.M. Hull and B.
Spence, Phys. Lett. {\bf B232} (1989), 204.}
\REF\isogauge {I. Jack,
D.R.T. Jones, N. Mohammedi and H. Osborn, Nucl. Phys.
{\bf B332} (1990), 359.}
\REF\hst{C.M. Hull and B.
Spence, Nucl. Phys.   {\bf B353}  (1991), 379.}
\REF\pq{C.M. Hull, G.
Papadopoulos and B. Spence, Nucl. Phys. {\bf B363} (1991) 593.
}
\REF\hullwitt {C. M. Hull and E. Witten, Phys. Lett. 160B (1985)
398.}
\REF\papa{G. Papadopoulos,  Phys. Lett. {\bf B238} (1990), 75.}
\REF\prep{C.M. Hull, in preparation.}
\REF\brooks {R. Brooks, F. Muhammad and S.J. Gates, Jr., Nucl. Phys.   {\bf
268} (1986) 599.}

\def\dm {\partial_{\mu}}
\def\dn {\partial_{\nu}}

\def\ffi {\phi^i}
\def\fj {\phi^j}
\def\fk {\phi^k}
\def\ix {\int\!\!d^2x  \sqrt{h} \;}

\def\ep {\epsilon^{\mu\nu}}
\def\liea { {\cal L}_a }
\def\hli {\hat { {\cal L}}_a }

\def\a {A^a_{\mu}}

\def\dpl {\partial_+}
\def\dpp {\partial_+}

\def\dmm{\partial_-}

\def\gij{{g_{ij}}}
\

\def\dpl{D }
 \def\nmm{\nabla_-}
\def\npp{\nabla_ +}
\def\thp{\theta }
\def\H{H_{ijk}}
\def\dd{{\cal D}}
\def\aa{{\cal A}}
\def\intxt{\int\!\!d^2x\,dt\,}
\def\del{\partial}
\def\xiai{\xi_a^i}

\def\intx{\int\!\!d^2x\,}
\def\bij{b_{ij}}

Non-linear sigma-models are important two-dimensional field theories and those
that are conformally invariant describe the propagation of a string in a curved
space-time   [\sen]. Gauging such sigma-models can give  a construction of new
conformal field theories, and gauged Wess-Zumino-Witten models provide a
lagrangian formulation of    the coset construction [\gauged] .
More recently, duality symmetry in string theory has been formulated in terms
of gauged sigma-models [\rocver],
and this has led to the proposal of non-abelian generalisatons of duality
symmetry [\queossa]. Supersymmetric gauged sigma-models have also been used to
construct (p,q) supersymmetric integrable models [\pots].
In [\hso,\isogauge,\hst], the gauging of  general non-linear sigma-models with
Wess-Zumino terms was shown to be possible  only if certain obstructions were
absent, and the
gauged action was given. The (p,q) supersymmetric generalisations of these
sigma-models were constructed in [\pq].
The purpose of this paper is to extend these results to the case of
sigma-models with fermionic terms, representing the interaction of strings with
background Yang-Mills fields $C_i^{AB}$, in addition to the metric $\gij$,
anti-symmetric tensor gauge field $\bij$ and dilaton $ \Phi$. In particular,
the (1,0) supersymmetric version of such terms describes the propagation of
heterotic strings in backgrounds with non-trivial  gauge fields [\hullwitt]
and the gauged version can be used to formulate the effect of duality
transformations on Yang-Mills fields [\prep].

The action for a bosonic  two dimensional sigma-model with Wess-Zumino term and
Fradkin-Tseytlin term is
$$ S_0 = {1\over {2 }} \ix  \left( g_{ij}\partial^{\mu}\ffi\dm\fj
 +\ep b_{ij}\dm\ffi\dn\fj + \Phi R \right)
\eqn\ones  $$
where the  $D$-dimensional target space $M$ has metric $g_{ij}(\phi)$,
coordinates
$\phi^i \  (i=1,...D)$  and torsion three-form $H$ given in terms of a
potential  $b_{ij}$ by
$H_{ijk}= {3\over2}\partial_{[i}b_{jk]} $.
It is invariant under the transformation
$$  \delta\ffi =   e\lambda^a\xi^i_a
\eqn\sixes  $$ where $\lambda^a$ $ (a=1,...n)$ are infinitesimal constant
parameters, $e$ is a constant and $\xi_a$
are a set of vector fields on $M$
  provided the Lie derivative with respect to $\xi_a$ of  $g_{ij}$, $\Phi$  and
$H_{ijk}$ vanish, which will be the case if
$$ \nabla_{(i}\xi_{j)a} = 0                          \eqn\sevens  $$
(where $\xi _{ia}= \gij \xi^j_a$)
$$\xi^i_a \partial_i \Phi=0
\eqn\eerwr$$
and
 $\xi^i_aH_{ijk}$ is an exact two-form, \ie\ there is some (globally
defined) set of  Lie-algebra-valued one-forms $v_a$, defined up to the addition
of a closed form, with components $v_{ia}$ such that
$$  \xi^i_aH_{ijk} = \partial_{[j}v_{k]a}    \eqn\thirteens     $$
Then $\xi_a^i$ are Killing vectors  which can be taken to generate  some
$n$-dimensional isometry group $G$
satisfying
$$ [\xi_a , \xi_b ] \equiv \liea\xi_b = {f_{ab}}^c\xi_c  \eqn\eights  $$
where
${f_{ab}}^c$ are the structure constants of ${G }$ and $\liea$ denotes the
Lie derivative with respect to $\xi_a$.

The vanishing of the Lie derivative of $H_{ijk}$ implies that
$$\liea b_{ij}=\partial _{[i} \Lambda_{j]a}
\eqn\vab$$
 for some $\Lambda_{ia}$ so that the variation of $b_{ij}$ can be cancelled by
an anti-symmetric tensor gauge transformation, or equivalently, the variation
of the $b$-term in \ones\ is a total derivative.  If  $\Lambda_{ia}=0$, then
the symmetry can be gauged by minimal coupling, \ie\ by replacing the
derivatives $\partial_\mu$ in \ones\ by gauge-covariant derivatives $D_\mu$,
where
$$  D_{\mu}\ffi  = \dm\ffi - e\a \xi^i_a           \eqn\tens   $$
and the  gauge field $\a$ transforms as
$$ \delta\a = \dm\lambda^a + e{f^a}_{bc}A^b_{\mu}\lambda^c   \eqn\nines  $$
In the   case in which   $\Lambda_{ia} \ne
0$,   minimal coupling is not sufficient and the gauging is as given in
[\hso,\isogauge,\hst]. The gauging can in principle be given in terms of the
$\Lambda_{ia}$ given in \vab, but these are not vector fields
in general since the $b_{ij}$ are not tensors but are connections, and it is
more convenient to work in terms of the covariant $v_{ia}$.
Gauging of the isometry symmetry \sixes\ is possible only if
[\hso,\isogauge,\hst] (i) the $v_{ia}$ can be chosen  to be equivariant, \ie\
chosen so that
$$  \liea v_{ib} = {f_{ab}}^cv_{ic}                 \eqn\liev  $$
and (ii) if
$$  c_{(ab)}
=0\eqn\hiso$$
where  $$c_{ab} = v_{ia}\xi^i_b ,
 \eqn\cis$$
If these two conditions  are satisfied, then the gauged action is
$$\eqalign{
S_G& =   {1\over2} \ix \left\{ g_{ij}\, D_{\mu}\ffi D^{\mu}\fj   + \Phi R
\right\}
\cr &
+{1 \over 2} \!\ix \ep
\left( b_{ij}\dm\ffi\dn\fj
+2e \a v_{ia}\dn\ffi
-e^2
c_{[ab]}(\phi) \a A^b_{\nu}
 \right) \cr}
\eqn\sgag   $$
This can be rewritten as [\hso,\hst]
$$\eqalign{
S _G &=   {1\over2} \ix    \left\{ g_{ij}\, D_{\mu}\ffi D^{\mu}\fj   + \Phi R
\right\}
\cr &
+\int_Y\,\left({1\over3}H_{ijk} D\ffi D\fj D\fk
+ {e\over2}v_{ia}
   D \ffi F^a \right)
\cr}
 \eqn\elevens   $$
where $Y$ is a three-manifold whose boundary is the world-sheet $X$
and the field strength two-form
is $F^a = dA^a - {1 \over 2} e f^a _{\ bc} A^b A^c$.
As usual, the fields $\ffi, A_\mu ^a$ on $X$ are extended to fields on $Y$ in
the second term in \elevens.
If \liev\ is satisfied but \hiso\ is not, then the action \sgag\ is not
gauge-invariant, but satisfies
$$\delta S = e^2 \ix\ep  c_{(ab)} \a(\dn\lambda^b)
                                                        \eqn\twentyfives  $$
which is proportional to the consistent  chiral anomaly in two dimensions.

Suppose now that one adds a fermionic term of the form [\hullwitt]
$$S_f= {i\over2} \ix \psi_A (\nabla_+ \psi)_A
\eqn\sf$$
where $\psi_A$ are chiral Majorana world-sheet spinor fields  that are also
sections of an $O(N)$  vector bundle over $M $ with connection $C_i^{AB}(\phi)
$, $C_i^{AB}=- C_i^{BA}$, and fibre metric $\delta_{AB}$, which is used to
raise and lower the $O(N)$  vector indices $A,B,\dots=1, \dots,N$.
The covariant derivative is
$$
(\nabla_\mu \psi )_A= \partial_\mu \psi_A - {1 \over 2}\omega_ \mu \psi_A
-\partial_\mu \phi ^i C_i^{AB} \psi_B
\eqn\covis$$
where $ \omega_ \mu= {1 \over 2} \epsilon_{ab}\omega_\mu^{ab}$,
$\omega_\mu^{ab}$
is  the world-sheet spin-connection and $\nabla_ \pm = e^\mu_\pm \nabla_ \mu
$ where $e^\mu_a$ $(a=\pm)$ are zweibeins, with $e^\mu_\pm=\pm
\epsilon^{\mu \nu}e_{\nu \pm}$.
This is formally invariant under the $O(N)$ gauge transformations
$$
\delta\psi_A=M_A^{\ B} \psi_B, \qquad \delta C_i = \partial_i M - [C_i, M]
\eqn\onsymm$$
with parameter $M_A^{\ B} (\phi)$.
Under an arbitrary variation of the fields,
the action \sf\ changes by
$$\delta S_f={i } \ix \left\{ ( \Delta \psi_A )(\nabla_+ \psi)_A
- {1 \over 2} \delta \phi ^i \partial_+ \phi ^j G_{ij}^{AB} \psi_A\psi_B
\right\}
\eqn\vact$$
where the field strength is
$$G_{ij}= \partial_i C_j -\partial_j C_i -[C_i,C_j]\eqn\gis$$
and the covariant variation is defined by
$$\Delta \psi_A= \delta \psi_A - \delta \phi ^i C_i^{AB} \psi_B\eqn\covar$$

The transformation \sixes\  for constant $\lambda$ will lead to a symmetry of
the action $S_f$ if the connection
$C_i$ is invariant up to a gauge transformation:
$$\liea C_i = \nabla_ i \kappa_a\eqn\conc$$
 for some  $\kappa_a^{AB}(\phi)$, as the variation of the action can then be
cancelled by
an $O(N)$ transformation of $\psi_A$, $\delta \psi_A = \lambda^a \kappa_a ^{AB}
\psi_B$.
This condition has been
discussed in [\papa], where particular attention is paid to global aspects.
It can be reformulated covariantly as follows. The
condition for there to be an isometry symmetry is that the field strength
satisfy
$$
\xi^i_a G_{ij}^{AB}= \nabla_i \mu_a^{AB}
\eqn\mucon$$
for some $  \mu_a^{AB}
(\phi)$.
This is equivalent to \conc\ with
$$\mu_a= \kappa_a- \xi^i_a C_i
\eqn\kais$$
The $\kappa_a$ are not $O(N)$-covariant (they transform as a connection), and
it is more convenient to work with
the   $  \mu_a^{AB}$ defined by \mucon,
which transform covariantly under $O(N)$ transformations
$$\delta \mu_a = [M, \mu_a] \eqn\mutrans$$
   just as for the Wess-Zumino term it was better to work with the
$v_{ia}$ rather than the $\Lambda_{ia}$.
If \mucon\ is satisfied, then the action \sf\ is invariant under the rigid
transformations given by
\sixes\ and
$$
\Delta \psi_A=  e\lambda^a \mu_a ^{AB} \psi_B
\eqn\dely$$

Under the transformations given by  \sixes,\dely,    with local $ \lambda(x)$,
the action $S_f$ varies by
$$\delta S_f=e \ix  \partial_+ \lambda^a J_{a-}
\eqn\vod$$
where $J_{a-}
$ is the Noether current
$$J_{a-}
={i \over 2}( \psi_A \mu_a ^{AB} \psi_B)
\eqn\noeth$$
This variation can be cancelled by adding the Noether coupling $-eA_+J_-$ to
obtain
$$\eqalign{S_g&= {i\over2} \ix \left( \psi_A (\nabla_+ \psi)_A
- eA^a_+ (\psi \mu_a \psi) \right)
\cr &={i\over2} \ix \psi_A (\dd _+ \psi)_A
\cr}
\eqn\sfg$$
where
$$\eqalign{ (\dd _+ \psi)_A&= (  \nabla_+ \psi)_A -eA_+^a \mu_a ^{AB} \psi_B
=\partial_+ \psi_A - \aa _+^{AB} \psi_B
, \cr
\aa_\mu ^{AB}&=eA_\mu^a \mu_a ^{AB} + C_i^{AB} \partial_\mu \phi ^i
\cr}
\eqn\aais$$
This action is then fully gauge-invariant provided the $\mu_a$ are equivariant,
\ie\ they satisfy
$$\hli \mu_b -[\mu_a, \mu_b]= {f_{ab}}^c \mu_c\eqn\mueq$$
where
$\hli$ is the gauge-covariant Lie derivative, which for tensors
$T_{ij...}^{AB}$
transforming according to the adjoint of $O(N)$ is given by
$$\hli T_{ij...}^{AB}= \liea T_{ij...}^{AB} - \xi^i_a [C_i, T_{ij...}]^{AB}
\eqn\covli$$
This is the analogue of the equivariance condition \liev.

To summarise, the action \sf\  is invariant under rigid isometries provided
that
the field-strength satisfies \mucon\ for some $\mu_a$, and this can be promoted
to a local symmetry provided the $\mu_a$ satisfy the equivariance condition
\mueq, in which case the gauged action is \sfg.
Note that \mucon\ is equivalent to the condition that
$$
\hli G_{ij} =[\mu_a, G_{ij}]
\eqn\gooo$$
for some $ \mu_a$, so that the gauge-covariant Lie derivative of the
field-strength vanishes up to a gauge transformation.  The action \sfg\ is also
invariant under the $O(N)$ transformations
\onsymm,\mutrans.

As an example, consider the case in which the vector bundle is the tangent
bundle and $C_i^{AB}$ is the torsion-free spin-connection, where $A,B$ are now
tangent space indices with respect to a target space vielbein $E_i^A$. Then
$G_{ij}^{AB}=R_{ijAB}$ where $R_{ijAB}$ is the curvature tensor and  \mucon\ is
automatically satisfied with
$\mu_a^{AB}$ proportional to $E^A_iE^B_j \nabla^i \xi_a^j$.

The gauge variation of the connection $\aa_+$ defined by \aais\ is
$$\delta \aa_+^{AB} = \partial_+ \lambda^{AB} - [\aa , \lambda]
^{AB} , \qquad
\lambda
^{AB}  \equiv \lambda^a \kappa_a ^{AB}
\eqn\delaa$$
so that the action \sfg\ is manifestly invariant under the transformations
\delaa\ and
$\delta \psi_A= -e \lambda^{AB}  \psi_B$, which is equivalent to \dely.
Quantum mechanically, the gauge symmetry of the chiral fermion action \sfg\ is
anomalous, with
the variation of the effective action proportional to $$
\Delta
 = \ix \aa_+^{AB} \partial_-\lambda ^{AB}
\eqn\erwrr$$
Adding a counterterm proportional to $\Tr ( \aa _+ \aa _-)$, the anomaly
becomes
proportional to $$
\Delta
 = \int \aa ^{AB} d \lambda ^{AB}
\eqn\erwrra$$
where $\aa ^{AB}= \aa ^{AB}_ \mu dx ^ \mu$. When rewritten in terms of $A^a$
and $ \lambda^a$, this contains $ d \phi$ and $ d \phi d \phi$ terms.

It is straightforward to extend these results to the (1,0) supersymmetric
sigma-model.
For a flat world-sheet with $h_{\mu \nu}= \eta_{\mu \nu}$, flat (1,0)
superspace has coordinates
$x^+, x^-$ and $\theta$ and   flat superspace derivative $D$
 with $D^2 = i\dpp$. The (1,0) supersymmetric generalisation of \ones\  is
given
by [\hullwitt]
$$   S = -i \Big[\intx d\thp \gij\dpl\ffi\dmm\fj + \intxt
d\thp\H\del_t\ffi\dpl\fj
              \dmm\fk \Big],                                 \eqn\six $$
where
$\phi$ is now a superfield, \ie\ a map from the (1,0) superspace into $M$.
The transformations \sixes, now involving superfields, are
rigid symmetries of the action \six\ provided that
the vector fields $\xiai$
satisfy the same conditions   as in the bosonic case. To
promote  these
rigid symmetries to local ones,  it is necessary to introduce a (1,0) super
Yang-Mills multiplet. This is described  by gauge superfields $A (x ,
\theta)$, $A_+(x ,
\theta)$, $A_-(x ,
\theta)$ which can be used to define
  gauge-covariant derivatives $\nabla,\nmm,\npp $.
These are taken to satisfy the constraints
[\brooks]
$$  [\nabla,\nabla] = 2i\npp, \quad [\nabla,\nmm] = W_,\quad [\npp,\nmm] = F_{
},
                                                                \eqn\seven $$
with all other
super-commutators equal to zero. In equation \seven, the field strength $W^a$
is an unconstrained
superfield while the Bianchi identities imply that $F   ^a$ is
proportional to $D W $.
Acting on sigma-model fields, these become (setting $e=-1$)
$$\nabla   \ffi= D  \ffi + A ^a \xi ^i_a (\phi),
\quad [\nabla,\nmm]\ffi = W^a \xi ^i_a (\phi)
\eqn\exex$$
etc.
Gauging is possible if and only if it is possible for the corresponding bosonic
model, in which case
the action for the gauged (1,0) sigma
model
is
$$ \eqalign {  S = -i  \intx d\thp & \Big[\gij\nabla\ffi\nmm\fj + \bij\dpl\ffi
                    \dmm\fj
                  \cr  & -  A^a u_{ia}
\dmm\ffi
        + A_-^a u_{ia} \dpl\ffi + A^a A^b_-c_{[ab]}\Big], \cr } \eqn\eighta $$

The (1,0) supersymmetric generalisation of \sf\ is [\hullwitt]
$$   S =   {1 \over 2} \intx d\thp \psi_A \tilde D \psi_A
   \eqn\ssf $$
where
$$\tilde D \psi_A      =   D \psi_A     -D \phi ^i C_i ^{AB} \psi_B
\eqn\dti$$
and $\psi_A $ are now fermionic superfields.
Again, this is invariant under rigid isometry symmetries if and only if the
connection $C_i(\phi)$ satisfies \mucon\ for some $\mu_a$, and the rigid
symmetry can be gauged  if $\mu_a $ is equivariant \mueq, in which case the
gauged action is the (1,0) supersymmetrisation of \sfg, given by
$$   S =  {1 \over 2} \intx d\thp \psi_A \hat D \psi_A
\eqn\sasf $$
where
$$\hat  D \psi_A      =  \tilde D \psi_A   +  A^a \mu_a ^{AB} \psi_B
\eqn\dtia$$

\refout
\bye